\documentclass{PoS}

\usepackage{amsmath,amsfonts,amssymb,epsfig,comment}

\def\twovec[#1,#2]{\left( \begin{array}{c} #1  \\ #2 \end{array} \right)}
\def\twomat[#1,#2][#3,#4]{\left( \begin{array}{cc} #1 & #2 \\ #3 & #4 \end{array} \right)}
\def\threemat[#1,#2,#3][#4,#5,#6][#7,#8,#9]{\left( \begin{array}{ccc} #1 & #2 & #3\\ #4 & #5 & #6 \\ #7 & #8 & #9 \end{array} \right)}

\def\Ms{M_{SUSY}} 
\def\MG{m_{3/2}}
\def\MP{M_{Pl}}

\def\mrm{\mathrm}

\title{Stripping a supermultiplet of all but one scalar: a microscopic view}

\ShortTitle{Stripping a supermultiplet of all but one scalar}

\author{\speaker{Karim Benakli}\\
        Sorbonne Universit\'e, UPMC Univ Paris 06, UMR 7589, LPTHE, F-75005, Paris, France \\
CNRS, UMR 7589, LPTHE, F-75005, Paris, France\\
        E-mail: \email{kbenakli@lpthe.jussieu.fr}}

\author{Yifan Chen\\
        Sorbonne Universit\'e, UPMC Univ Paris 06, UMR 7589, LPTHE, F-75005, Paris, France \\
CNRS, UMR 7589, LPTHE, F-75005, Paris, France\\
       E-mail: \email{yifan.chen@lpthe.jussieu.fr}}

\abstract{ We explain how a single scalar degree of freedom can be obtained from  projecting out all the other components of a chiral superfield using simple operators in the microscopic theory.  We use the Fayet-Iliopoulos model as an example of the origin for the necessary supersymmetry breaking. We also comment on one peculiar aspect where non-linear realisation can be useful: the freeze-in scenario for gravitino dark matter.}

\FullConference{Corfu Summer Institute 2017 'School and Workshops on Elementary Particle Physics and Gravity'\\
		2-28 September 2017\\
		Corfu, Greece}

\begin{document}

\section{Introduction}
\label{SEC:INTRO}

Supersymmetric field  theories are important playgrounds for theoretical physicists. Supersymmetry allows better control of some quantum corrections and therefore some symmetries beyond classical level. However, motivated by the study of our Universe, it is essential to investigate non-supersymmetric theories and possibly those which arise when supersymmetry is hidden (broken) at low energies. Two questions arise then which are of interest here: what is peculiar at low energies to theories that are only supersymmetric in the ultra-violet? which of the tools developed  for study these theories in the ultra-violet are still useful in the infra-red?  As an answer for the first, there is the fact that when the breaking of supersymmetry is spontaneous, there is a goldstone fermion, the goldstino. In supergravity, it is absorbed by the gravitino making its longitudinal mode dynamical. In the global supersymmetric limit, it is massless, thus present in the low energy theory. The simplest corresponding  low energy effective theory is described by the Volkov-Akulov action  \cite{Volkov:1973ix}. One aspect of  the second question is the possible use of superspace and superfields in the low energy non-supersymmetric effective theory.  Superfields make the construction of Lagrangians easier. Starting from  \cite{Rocek:1978nb}, many works discussed the use of superfields  for non-supersymmetric theories. We shall review here the results of \cite{Benakli:2017yar}.


\section{Integrating out without explicit integration}
\label{SEC:INTRO}

Let us first start with the simplest case where supersymmetry breaking gives rise to a large mass to the scalar component $\phi$ of a supermultiplet. Integrating out these states amounts to express it as a function of $\psi$ and $F$,  the fermionic  and auxiliary field components of the chiral supermultiplet, respectively. We can use the supersymmetry transformations:
\begin{eqnarray}  
\delta_\epsilon\phi (\psi, F) &&= \frac{\partial\phi}{\partial\psi_\alpha} \delta_\epsilon\psi_\alpha +  \frac{\partial\phi}{\partial F} \delta_\epsilon F \nonumber \\
\epsilon\psi &&= \frac{\partial\phi}{\partial\psi_\alpha} [-i(\sigma^\mu\overline{\epsilon})_\alpha \partial_\mu\phi + \epsilon_\alpha F] - \frac{\partial\phi}{\partial F} (i\overline{\epsilon}\overline{\sigma}^\mu\partial_\mu\psi).
\end{eqnarray}
One can solve the partial differential equation by taking:
\begin{equation} \phi = \frac{\psi\psi}{2F}.\label{sgoldstino}\end{equation}
Note that the same result can be obtained by writing down the Lagrangian and solving the corresponding equations of motion. Putting this result back in the chiral multiplet:
\begin{equation} X_{NL} = \frac{\psi\psi}{2F} + \sqrt{2}\theta\psi + \theta\theta F,\end{equation}
one sees that it is nilpotent:
\begin{equation} X_{NL}^2 = 0.\end{equation}
Without supersymmetric covariant derivatives for this superfield, the corresponding Lagrangiancan be written as:
\begin{eqnarray}\mathcal{L}_X &&= \int d^4\theta\overline{X_{NL}}X_{NL} + (\int d^2\theta f X_{NL} + h.c.)\nonumber\\
&&= i\overline{\psi}\overline{\sigma}^\mu\partial_\mu\psi - \partial^\mu (\frac{\overline{\psi\psi}}{2\overline{F}})\partial_\mu(\frac{\psi\psi}{2F}) + \overline{F}F + fF + \overline{f} \overline{F},\label{Eq:L_X}\end{eqnarray}
where f is a constant with mass dimension 2. Because of the nilpotency constraint, there are no higher order terms in $X_{NL}$.

We turn now to the case of the simplest gauge multiplet, a $U(1)$ vector multiplet. For simplicity, we choose the Wess-Zumino gauge. As fixing a super-gauge breaks supersymmetry, a supersymmetry transformation should be followed by a new super-gauge transformation in order to
go back to the Wess-Zumino gauge.  The transformation is:
\begin{eqnarray} 
\delta_{\Lambda} V &&= i(\overline{\Lambda} - \Lambda),\nonumber\\
\Lambda(y) &&= \frac{i}{\sqrt{2}}\theta\sigma^\mu\overline{\epsilon}A_\mu -\theta\theta\frac{i}{\sqrt{2}}\overline{\epsilon\lambda},
\label{gaugetransformationbacktoWZ}
\end{eqnarray}
Combing supersymmetry  and super-gauge transformations with parameters $\epsilon$  and ${\Lambda}$, respectively, gives:
\begin{eqnarray} \sqrt{2}\delta_{\epsilon + {\Lambda}} A_\mu &&= \epsilon\sigma_\mu\overline{\lambda} + \lambda\sigma_\mu\overline{\epsilon} ,\nonumber \\
\sqrt{2}\delta_{\epsilon + {\Lambda}}\lambda_\alpha &&= \frac{i}{2} (\sigma^\mu\overline{\sigma}^\nu\epsilon)_\alpha F_{\mu\nu} + \epsilon_\alpha D ,\nonumber\\
\sqrt{2}\delta_{\epsilon + {\Lambda}} D &&=  -i\overline{\epsilon}\overline{\sigma}^\mu\partial_\mu\lambda + i\partial_\mu\overline{\lambda}\overline{\sigma}^\mu\epsilon .
\end{eqnarray}
We are interested by the simplest low energy effective action where the gauge group and supersymmetry are broken so that we can integrate out all fields including $A_\mu$ and remain only with the goldstino. For this purpose, we work in the superunitary gauge. This is an unusual gauge at the level of superspace but it appears quite convenient for the problem of interest here. When integrated out, the component $A_\mu$ has to appear as a function of the other fields, and in the chosen gauge, at first order of $\lambda$, $\overline{\lambda}$ and D. Supersymmetric transformations:
\begin{equation} \delta_\epsilon A_\mu = \frac{\partial A_\mu}{\partial \lambda_\alpha} [\frac{i}{2\sqrt{2}} (\sigma^\mu\overline{\sigma}^\nu\epsilon)_\alpha F_{\mu\nu} + \frac{1}{\sqrt{2}}\epsilon_\alpha D] + h.c. + \frac{\partial A_\mu}{\partial D} \frac{i}{\sqrt{2}} (-\overline{\epsilon}\overline{\sigma}^\mu\partial_\mu\lambda + \partial_\mu\overline{\lambda}\overline{\sigma}^\mu\epsilon).\end{equation}
close if:
\begin{equation} A_\mu = \frac{\lambda\sigma_\mu\overline{\lambda}}{D}.\label{heavygaugeboson}\end{equation}
Note that gauge invariance is not manifest as the gauge symmetry is broken and we are working in the unitary gauge. In a different gauge,  the would-be Goldstone boson $a$ should appear as
$$
A_\mu - \frac{1}{m_{A}} \partial_\mu a= \frac{\lambda\sigma_\mu\overline{\lambda}}{D},
$$
which goes to the above expression when $m_A \rightarrow \infty$. 
The effective Lagrangian for this $U(1)$ theory will then take the formn:
\begin{eqnarray} \mathcal{L}_V &&=  \int d^2\theta \frac{1}{4} W_{NL}W_{NL} + h.c. + \int d^4\theta2\xi V\nonumber\\
&&=  i\overline{\lambda}\overline{\sigma}^\mu\partial_\mu\lambda -\frac{1}{4}[\partial_\mu(\frac{\lambda\sigma_\nu\overline{\lambda}}{D}) - \partial_\nu(\frac{\lambda\sigma_\mu\overline{\lambda}}{D})][\partial^\mu(\frac{\lambda\sigma^\nu\overline{\lambda}}{D}) - \partial^\nu(\frac{\lambda\sigma^\mu\overline{\lambda}}{D})]+ \frac{1}{2}D^2 + \xi D\nonumber,\\\end{eqnarray}
which can be shown to be equivalent to eq. (\ref{Eq:L_X}) when $\xi = \sqrt{2}f$.

The main lesson to learn from the above discussion is the fact that it is possible to describe the low energy effective action of theories with spontaneously broken supersymmetry using a superfield built only from the goldstino. The part of the discussion on the Fayet-Iliopoulos model is new, though straightforward, and its interest, or value, steams from possible applications as we shall discuss below. We will first summarize some old and useful properties of the nilpotent superfields.

\section{Chiral and vector nilpotent superfields}
\label{SEC:INTRO}

Early examples  \cite{Rocek:1978nb,Casalbuoni:1988xh} introduce a superfield $X_{NL}$ which satisfies: 
\begin{eqnarray}
X_{NL} ^2=0
\label{squareX}
\end{eqnarray}
This constraint eliminates a component, and it is easily seen to be the scalar component, often denoted nowadays as the sgoldstino. Fixing the scale of supersymmetry breaking, the auxilliary component $F_X$ of  $X_{NL}$ can be achieved by imposing an additional constraint \cite{Rocek:1978nb} :
\begin{eqnarray}
X_{NL} \overline{D}^2\overline{X_{NL} } \propto X_{NL} 
\label{Ffixed}
\end{eqnarray}
fixes the scale of supersymmetry breaking, the F-term $F_X$ in $X_{NL}$ \cite{Rocek:1978nb}. Then:
\begin{eqnarray}
D_\alpha X_{NL} | = \sqrt{2}\tilde{\lambda}_\alpha + \cdots ; \qquad X_{NL} |  = \frac{\tilde{\lambda}_\alpha\tilde{\lambda}^\alpha}{2F_X} +\cdots
\label{goldstinoX}
\end{eqnarray}
and the goldstino is the only degree of freedom in the superfield.

Another important result is the fact that for the violation of  the conservation of the Ferrara-Zumino supercurrent $\mathcal{J}_{\alpha \dot{\alpha}}$:
\begin{eqnarray}
\overline{D}^ {\dot{\alpha}} \mathcal{J}_{\alpha \dot{\alpha}} = D_\alpha X
\end{eqnarray}
the superfield $X$ flows in the infrared to $X_{NL}$, i.e. $X \rightarrow X_{NL} $ \cite{Komargodski:2009rz}. Other ways to embed the goldstino in a constrained superfield can be found in \cite{Lindstrom:1979kq,Samuel:1982uh,Ivanov:1978mx,Ivanov:1982bpa,Samuel:1982uh}.

As we are interested in a $U(1)$ gauge theory, we want to consider goldstinos that arise from vector multiplet $V_{NL} $ components instead of chiral multiplet $X_{NL}$.  Appropriate constraints in this case can be taken to be:
\begin{eqnarray} V_{NL} = {V_{NL} }^\dagger \label{reality} \end{eqnarray}
\begin{eqnarray} V_{NL} ^2= 0 \label{nilpotency} \end{eqnarray}
\begin{eqnarray} V_{NL}  \propto V_{NL}  (D^\alpha \overline{D}^2 D_\alpha + \overline{D}^{\dot{\alpha}} {D}^2 \overline{D}_{\dot{\alpha}}) V_{NL}   \label{Dfixed} \end{eqnarray}
where the goldstino  is embedded as the lowest component of
\begin{eqnarray} W_{NL \alpha} = -\frac {1}{4}  \overline{D}^2 D_\alpha  V_{NL}  = \tilde{\lambda}_\alpha + \cdots   \label{Wgoldstino} \end{eqnarray}
A solution for these constraints is
\begin{eqnarray}
V_{NL}  = \frac{\overline{X_{NL} }X_{NL}}{F_X^2} 
\end{eqnarray}

Note that  $V_{NL} $ can be used either for a true D-term breaking model  or to parametrise the effects of an F-term breaking as done in \cite{Bandos:2016xyu,Cribiori:2017ngp, Buchbinder:2017qls,GarciadelMoral:2017vnz}. Here the condition (\ref{Dfixed}) appears as a consequence of (\ref{Ffixed}).

Finally, let us note that, as expected for the case of a low energy Lagrangian of a single goldstino,  it was shown explicitly in  \cite{Lindstrom:1979kq,Samuel:1982uh,Ivanov:1982bpa,Luo:2009ib,Liu:2010sk,Kuzenko:2010ef}  that the corresponding Lagrangian is the Volkov-Akulov one  up to non-trivial field redefinitions.

\section{Use of nilpotent superfields}
\label{SEC:INTRO}

The nilpotent superfields are useful to ease the extraction of the couplings of goldstino to matter fields. This can be shown in simple low energy effective theories that do not include complicated sectors of supersymmetry breaking . More precisely, constraints  allow to project out heavy components, to perform their integration out as functions of the remaining light modes, without explicitly going through the ultraviolet (UV) microscopic Lagrangian. 

\subsection{Stripping superfields of their fermionic components}

We have seen above how it is easy to get rid of a scalar component. What about fermionic components? or even more components?

May be the most interesting case is to get rid of all but one scalar degree of fredom. It was shown in \cite{Komargodski:2009rz} that this can be done through the constaint:
\begin{equation} 
X_{NL} (\mathcal{A} + \overline{\mathcal{A}}) = 0, 
\label{constrintIntro}
\end{equation}
which allows to integrate out all components of the chiral multiplet $\mathcal{A}$ leaving only a pseudo-scalar degree of freedom propagating. This is useful for models of goldstone bosons \cite{Komargodski:2009rz} and applications for example to  infation \cite{Ferrara:2015tyn, Dalianis:2017okk}.

There are many reasons why one may want to understand how (\ref{constrintIntro}) arises from a spontaneous breaking of a linearly realised supersymmetry theory in the UV. For example, one needs to check that this is not the result of a pathological way to break supersymmetry, also what kind of additional sectors are required to be present in the fundamental theory?.

Clearly, simplest constraints eliminate only one degree of freedom. Therefore, this constraint has to be a combination of simpler ones. This was shown explicitly in \cite{DallAgata:2016syy}. The authors found that (\ref{constrintIntro}) is equivalent to imposing three  constraints:
\begin{equation} \overline{X_{NL}}X_{NL} (\mathcal{A}+ \overline{\mathcal{A}}) = 0, \end{equation}
\begin{equation} \overline{X_{NL}}X_{NL}\overline{D_{\dot{\alpha}}\mathcal{A}} = 0, \end{equation}
\begin{equation} \overline{X_{NL}}X_{NL}\overline{D^2 \mathcal{A}} = 0, \end{equation}
which eliminates the heavy real scalar, the fermion and the auxiliary field, respectively.  

However, writing a UV Lagrangian for these constraints is not straightforward. In \cite{DallAgata:2016syy}, it was proposed to add  these three operators to the Lagrangian with Lagrange multipliers that are then sent to infinity. This is certainly the simplest possibility but it has  the inconvenience of introducing higher (covariant) derivative terms. We shall provide below an alternative with a single  operator that leads  to the constraint (\ref{constrintIntro}) based on the presence of a non-vanishing D-term that breaks supersymmetry.

\subsection{A single operator for the minimal constrained superfield}

 

Let us  consider the gauge invariant interaction between matter represented by a chiral superfield $\mathcal{A}^a$, the field strength of a gauge superfield $W^a_\alpha$ and the nilpotent goldstino superfield discussed above and represented by $W_{NL}^\alpha$:
\begin{eqnarray}
 -\frac{m_D}{f}\int d^2\theta W_{NL}^\alpha W^a_\alpha\mathcal{A}^a 
= -\frac{m_D}{4\sqrt{2}f^2}\int d^2\theta\overline{D}^2D^\alpha(\overline{X_{NL}}X_{NL})W^a_\alpha\mathcal{A}^a.
\label{LDG}
\end{eqnarray}
This the well known operator that is introduced to give Dirac masses for gauginos. Precisely, expanding in components, one identifies that the fermions $\lambda^a$ from $W^a$ and $\chi^a$ from $\mathcal{A}^a$ combine to make a Dirac fermion of mass $m_D$.  Solving the  $D$-term equations of motion in $W^a_\alpha$ shows that the real part of the scalar in $\mathcal{A}^a$ has a mass $2|m_D|$. These states decouple from the low energy theory as we take $m_D \rightarrow \infty$. The remaining light propagating states are: the goldstino, the gauge boson in $W^a_\alpha$ and the imaginary part of the scalar in $\mathcal{A}^a$.  The effective goldstino interactions for the case of Dirac gauginos of that operator have been studied in \cite{Goodsell:2014dia}. Here, we want to describe this decoupling by constraint equations.

The equation of motion to the $\mathcal{A}^a$ reads:
\begin{equation} 
\overline{D}^2D^{\alpha}(\overline{X_{NL}}X_{NL})W^a_{\alpha} = 0.
\end{equation}
Multiplied by $\overline{X_{NL}}X_{NL}D_\beta$ to the l.h.s., using the the nilpotency $XD^\alpha X = 0$ and the non-zero property of  $D\overline{D}^2D(\overline{X_{NL}}X_{NL})$ this gives:
\begin{equation} 
\overline{X_{NL}}X_{NL}W_\alpha^a = 0.
\end{equation}
This constraint projects out the gaugino in $W^a_\alpha$, as the result of the large Dirac mass.

Next, using the equation of motion of $W^a_\alpha$ leads to:
\begin{equation} 
D_\alpha\overline{D}^2D^\alpha(\overline{X_{NL}}X_{NL}) (\mathcal{A}^a + \overline{\mathcal{A}^a}) - [\overline{D}^2D^\alpha(\overline{X_{NL}}X_{NL})D_\alpha\mathcal{A}^a + h.c.] = 0.
\label{eomW}
\end{equation}
Again, multiplication by $\overline{X_{NL}}X_{NL}$ to the l.h.s., and getting rid of the second term using the nilpotency of $X_{NL}$ gives the constraint:
\begin{equation}
\overline{X_{NL}}X_{NL} (\mathcal{A}^a + \overline{\mathcal{A}^a}) = 0.
\label{yougotit}
\end{equation}
This constraint projects out the real part of the scalar.

Then, we  plug into the l.h.s. of Eq.(\ref{eomW}) $\overline{X_{NL}}X_{NL}D_\beta$ leading to:
\begin{equation}\overline{X_{NL}}X_{NL}D_\alpha\mathcal{A}^a = 0,\end{equation}
This constraint projects out the fermion $\chi^a$. 

Finally, we can write the constraint that eliminates the auxiliary field as:
\begin{equation}
\overline{X_{NL}}X_{NL}D^2\mathcal{A}^a = 0.
\end{equation}

Therefore, we have achieved the construction of the constraint of the previous subsection with a simple operator.

\subsection{Constrained superfield for axion}

A simple application of the previous operator is the case of $U(1)$. The operator (\ref{LDG}) can be used to describe an axion superfield coupled to the kinetic mixing between two different U(1) vector multiplets.  The corresponding constraints can lead to massive saxion and axino while leaving a light axion, as desired.

Writing the operator as:
\begin{equation} 
\mathcal{L}_{axion} = \frac{1}{f_A} \int d^2\theta W_{NL}^\alpha W_\alpha^{U(1)}A
\end{equation}
one can identify $A$ as the axion superfield, $W_\alpha^{U(1)}$ ias an abelian vector superfield while $f_A$ plays the role of the decay constant for the axion. Precisely, this has the form of (\ref{LDG}) therefore the axion coupling operator is exactly the same as the mass operator for the $U(1)$ Dirac gaugino when the singlet chiral superfield $\Sigma^1$ is the axion superfield. From this we can extract a useful relation between the supersymmetry breaking mediation and the axion symmetry breaking scales:
\begin{equation} 
m_D \sim \frac{f}{\Lambda} \sim \frac{f}{f_A} \rightarrow \Lambda \sim f_A.
\end{equation}

As required from the construction of an axion model, we have kept  the CP-odd scalar $a$ remains. It inherits a coupling to the goldstino through the  kinetic mixing:
\begin{equation}
\frac{a}{f_A}\epsilon^{\mu\nu\rho\sigma}F^{U(1)}_{\mu\nu}F^{NL}_{\rho\sigma},
\end{equation}
 between the $U(1)$ and a Fayet-Iliopoulos type $U(1)$.

\section{Splitting the Higgs supermultiplets }

An application of the operators introduced above of is splitting multiplets to extract the Standard Model Higgs $SU(2)$ doublet from a  minimal supersymmetric electroweak sector with two Higgs doublet superfields.  This implies projecting out the fermionic higgsinos and keeping as light field only one linear combination of the two Higgs doublets. This can be obtainedd using two operators, one to generate  a $\mu$-like term for the fermions,  resulting also in diagonal soft-terms and  a second one for the $B\mu$-term.

First, consider two chiral doublets $H_{1,2}$ with opposite $U(1)$ charges. Following \cite{Nelson:2015cea},  we introduce the operator:
\begin{eqnarray} 
\mathcal{O}_{H _{jj}}&&= \frac{a_{jj} m_{H} }{8\sqrt{2}f}\int d^2\theta\overline{D}^2(D^\alpha V_{NL}D_{\alpha}H_i)H_j\nonumber\\
&&= -\frac{a_{jj} m_H}{2\sqrt{2}f}\int d^2\theta W_{NL}^\alpha (D_{\alpha}H_i) H_j + ...,
\end{eqnarray}
for $i,j=1,2$.  Here ... represent  terms that do not contribute to the superpotential. A different non-linear realisation of the MSSM  can be found in \cite{Antoniadis:2010hs}. 

The above operator gives rise to a Dirac mass $\frac{1}{2}a_{ii} m_H \tilde{H_1} \tilde{H_2} $ for the fermionic modes $\tilde{H_1} \tilde{H_2} $. It also gives a mass $|a_{ii}|^2 |m_H|^2 |{H_i}|^2$ for the complex scalar in $H_i$ but  leaves the scalar in $H_j$ massless.  For a large scale supersymmetry breaking, only the scalar component in $H_j$ survives at low energy. The corresponding constrains read:
\begin{eqnarray} 
\overline{X_{NL}}X_{NL} D_\alpha H_i &&= 0;\nonumber\\
\overline{X_{NL}}X_{NL} D_\alpha H_j &&= 0;\nonumber\\
\overline{X_{NL}}X_{NL} H_j &&= 0\label{1HD} \ .
\end{eqnarray}

Next, introduce the additional operator:
\begin{eqnarray} 
\mathcal{O}_{H _{12}} &&= -\frac{a^2_{12} m_{H}^2 }{2f^2}\int d^2\theta  W_{NL} W_{NL} H_1 H_2 \ .
\label{DoubleW}
\end{eqnarray}
This generates an off-diagonal mass for the scalars $a^2_{12} m_{H}^2 H_1 H_2$ leading to a mass matrix:
\begin{equation}
m_H^2
  \begin{pmatrix}
    a_{11}^2 & a_{12}^2 \\
    a_{12}^2 & a_{22}^2
  \end{pmatrix}\end{equation}
 Taking $a_{ij}$ to infinity and  $a_{11} a_{22}-a^2_{12} \simeq 0$ allows to keep one state only light. The equation of motion of either $H_i$ reads:
\begin{equation} 
\frac{1}{4}\overline{D}^2\overline{H_i} = -\frac{a_{ii}m_H}{2\sqrt{2}f}W^\alpha_{NL}D_\alpha H_j - \frac{a_{jj}m_H}{2\sqrt{2}f}D_\alpha(W^\alpha_{NL}H_j) - \frac{a_{12}^2m_H^2}{2f^2}W_{NL}W_{NL}H_j, 
\label{eomHiggs}
\end{equation}
where the l.h.s from the kinetic term is included since the F-term of $H_i$ contributes to the mass term of $h_j$. Projecting by $X_{NL}\overline{X_{NL}}$, $X_{NL}\overline{X_{NL}}D_\beta$ and $X_{NL}\overline{X_{NL}}D^2$ respectively onto eq. (\ref{eomHiggs}):
\begin{eqnarray} 
\frac{1}{4}X_{NL}\overline{X_{NL}} \overline{D}^2\overline{H_i} &&= \frac{a_{jj}m_H}{2\sqrt{2}f} X_{NL}\overline{X_{NL}}D^\alpha W_{NL\alpha}H_j;\label{projectXX}\\
  0 &&= X_{NL}\overline{X_{NL}}D^\alpha W_{NL\alpha} (a_{ii} + a_{jj}) D_\beta H_j;\label{projectXXD}\\
  0 &&= \frac{a_{ii}m_H}{2\sqrt{2}f}X_{NL}\overline{X_{NL}}D^\alpha W_{NL\alpha}D^2 H_j + \frac{a_{12}^2m_H^2}{2f^2}X_{NL}\overline{X_{NL}}(D^\alpha W_{NL\alpha})^2H_j.\label{projectXXDD}\nonumber\\
  \end{eqnarray}
  Eq. (\ref{projectXXD}) allows to project the Higgsino through the constraint:
  \begin{equation} 
  \overline{X_{NL}}X_{NL} D_\alpha H_1 = \overline{X_{NL}}X_{NL} D_\alpha H_2 = 0.
  \end{equation}
Pluging eq. (\ref{projectXX}) into eq. (\ref{projectXXDD}) provides the constraint for projecting out the heavy higgs:
\begin{equation} \overline{X_{NL}}X_{NL} (a_{12}^2 H_j + a_{ii}^2\overline{H_i}) = 0.\end{equation}
Using   $a_{11} a_{22}-a^2_{12} =0$, this becomes equivalent to
\begin{equation} 
\overline{X_{NL}}X_{NL} (a_{11} H_1 + a_{22}\overline{H_2}) = 0. 
\end{equation}

\section{Nilpotent goldstino superfield from FI model}

The original Fayet-Iliopoulos (FI) model \cite{Fayet:1974jb} provides the simplest example of microscopic theory for $D$-term supersymmetry breaking.  We will show how $V$ flows in the infrared to $V_{NL}  \propto \overline{X_{NL} }X_{NL} $ where $X_{NL}$ is the goldstino nilpotent superfield  when both supersymmetry and the gauge symmetry are spontaneously broken leaving in the infrared only the massless goldstino. To illustrate how this arises, we will show this result in three different ways: from integrating out heavy modes within the Lagrangian in components fields, identification of the nilpotent superfield in the Ferrara-Zumino supercurrent equation and from integrating out the heavy modes through the superfield equations in the super-unitary gauge.

The Fayet-Ilioupous (FI) model contains two chiral superfields $\Phi_\pm (y, \theta, {\bar{\theta}}) = \phi_\pm (y)+ \sqrt{2} \theta \psi_\pm (y) + \theta \theta F_\pm (y)$, $y^\mu \equiv x^\mu - i \theta \sigma^\mu {\bar{\theta}}$. The Lagrangian is given by:  
\begin{eqnarray}
\int d^2\theta (\frac{1}{4}W^\alpha W_\alpha + m \Phi_+ \Phi_-) + h.c. + \int d^4\theta \left[  \overline{\Phi_+} e^{2gV} \Phi_+ + \overline{\Phi_-} e^{-2gV} \Phi_- + 2 \xi V \right] \, .
\end{eqnarray}
with $\xi $ the FI term associated to a $U(1)$ gauge field.
Solving for the auxiliary D-term, we obtain the potential 
\begin{eqnarray}
{\mathcal{L}} \supset& -m^2 (|\phi_+|^2 + |\phi_-|^2) - \frac{1}{2}( \xi + g |\phi_+|^2 - g |\phi_-|^2)^2 \, .
\end{eqnarray}
We are interested by the case $\xi g > m^2$ where both the $U(1)$ symmetry and supersymmetry are broken. Writing $\phi_- =\frac{1}{\sqrt{2}} (v + h +ia)$,  the solution to the equations of motion reads:
\begin{eqnarray}
\frac{g^2 v^2}{2} =& \xi g - m^2  \nonumber\\
D= - \xi + \frac{gv^2}{2} = - \frac{m^2}{g}  \, ,& \qquad F_+^* = -\frac{mv}{\sqrt{2}}\nonumber\\
|F_+|^2 + \frac{1}{2} D^2 =& \frac{m^2}{2 g^2} ( m^2 + g^2 v^2)  \ .
\end{eqnarray}

As a result, two spinors $\psi_-$ and $\tilde{\psi}$ combine into a Dirac fermion of mass $\sqrt{m^2 + g^2v^2}$, one vector $v_\mu$ and the real scalar $h$ get a mass $gv$, one complex scalar field $\phi_+$ has a mass $\sqrt{2m^2}$ and one fermion $\tilde{\lambda}$ remains massless, the goldstino. The fermionic mass eigenstates  can be obtained from the original fields through the re-definition: 
\begin{eqnarray}
(m \psi_+ - gv \lambda) \psi_- \propto& \tilde{\psi} \psi_- \nonumber\\
\rightarrow \twovec[\tilde{\psi},\tilde{\lambda}] =& \frac{1}{\sqrt{m^2 + g^2v^2}} \twomat[m,-gv][gv,m] \twovec[\psi_+,\lambda]  \nonumber\\
\twovec[\psi_+,\lambda]  =& \frac{1}{\sqrt{m^2 + g^2v^2}} \twomat[m,gv][-gv,m] \twovec[\tilde{\psi},\tilde{\lambda}].
\label{fermionmasseigenstates}\end{eqnarray}

\subsection {Integrating out in components}

We  integrate out all of the massive fields, and only the massless fermion $\tilde{\lambda}$ remains. Using the equations of motion, we get:
\begin{eqnarray}
A_\mu =& -\frac{g}{m^2 + g^2 v^2} \tilde{\lambda} \sigma^\mu \bar{\tilde{\lambda}} + ... \nonumber\\
\phi_+ =& - \frac{ g^2 v }{\sqrt{2} m(m^2 + g^2 v^2)}   \tilde{\lambda}\tilde{\lambda} + \mathcal{O}(\tilde{\lambda}^4) \nonumber\\
\psi_- =& - \frac{g^3 v}{m(m^2 + g^2 v^2)^{3/2}} \overline{\tilde{\lambda}}\overline{\tilde{\lambda}} \tilde{\lambda} + ... \nonumber\\
\tilde{\psi} =& -\frac{g^3 v}{m(m^2 + g^2 v^2)^{2}}  i \sigma^\mu \partial_\mu (\tilde{\lambda}\tilde{\lambda} \overline{ \tilde{\lambda}}) + ...
\end{eqnarray}
The imaginary part $a$ of $\phi_-$  have been taken be $h=a=0$ as we are working in the unitary gauge.

We can therefore express $\lambda$ as a function of $\tilde{\lambda}$
\begin{eqnarray}
\lambda = \frac{gv}{\sqrt{m^2 + g^2 v^2}} \bigg[ \tilde{\lambda} + \frac{g^2 }{(m^2 + g^2 v^2)^{2}}  i \sigma^\mu \partial_\mu [(\tilde{\lambda}\tilde{\lambda}) \overline{ \tilde{\lambda}}] + ... \bigg]
\end{eqnarray}
and the original chiral and  vector multiplets become:
\begin{eqnarray}
 \Phi_{+} (\phi_+, \psi_+, F_+) &&\overset{IR}{ \longrightarrow}  \frac{gv}{\sqrt{m^2 + g^2v^2}}\Phi_+ (\frac{\tilde{\lambda}\tilde{\lambda}}{2\tilde{f}}, \tilde{\lambda}, \tilde{f}) \label{Eq:Phi1component}\\
V (\lambda, v^\mu, D) && \overset{IR}{ \longrightarrow}  \frac{m}{\sqrt{m^2 + g^2v^2}}V (\tilde{\lambda}, \frac{\tilde{\lambda}\sigma^\mu\overline{\tilde{\lambda}}}{\sqrt{2}\tilde{f}}, \sqrt{2}\tilde{f} ).\label{Eq:Vcomponent}\end{eqnarray}
This is exactly what was expected from (\ref{sgoldstino}) and (\ref{heavygaugeboson}), showing that corresponding Lagrangian can be mapped to the Volkov-Akulov action. 

\subsection {Integrating out in superspace}



We work in the super-unitary gauge where the chiral superfield $\Phi_-$ is "eaten" by the gauge superfield. The Lagrangian reads:
\begin{eqnarray}
 \mathcal{L}_{SU} = \int d^2 \theta ( \frac{1}{4} W^\alpha W_\alpha +\frac{1}{\sqrt{2}} mv\Phi_+ ) + h.c. +\int d^4\theta ( \overline{\Phi_+} e^{2gV} \Phi_+ + \frac{1}{2} v^2e^{-2gV} + 2 \xi V ). \label{Eq:FILagrangeSU}
\end{eqnarray}
Working with superfields, we have to integrate out  via the equations of motion one of the superfields \emph{entirely} while leaving the other light. This can be done only if there is a hierarchy of masses. We need then to consider the two distinct limits $m^2\ll g^2v^2$ and $m^2\gg g^2v^2$ separately.  From the component calculation, we saw that in the first limit the Goldstino is mainly $\psi_+ \supset \Phi_+$, while in the second it is mainly the gaugino. Unsurprisingly, we will show that what remains in the spectrum in each limit is the corresponding superfield.

\subsubsection*{Case $\mathbf{m^2\ll g^2v^2}$:}

In this limit,  the goldstino is dominated by $\psi_+$. The equation of motion for $V$ is:
\begin{eqnarray}
0 =& \frac{1}{8} (D^\alpha \overline{D}^2 D_\alpha + h.c. ) V + 2 g \overline{\Phi_+} e^{2gV} \Phi_+ - g v^2e^{-2gV} + 2 \xi .
\end{eqnarray}
We write
\begin{eqnarray}
V =& \theta^4 \frac{1}{2} (-\xi + \frac{gv^2}{2}) + \hat{V} \equiv \theta^4 \frac{1}{2} \delta + \hat{V} \nonumber\\
W_\alpha =& \theta_\alpha \delta + \hat{W}_\alpha
\end{eqnarray}
and substituted back into the action, this gives
\begin{eqnarray}
\mathcal{L}_{SU} 
=& \int d^2 \theta ( \frac{1}{4} \hat{W}^\alpha \hat{W}_\alpha +\frac{1}{\sqrt{2}} mv\Phi_+ ) + h.c. +\int d^4\theta ( \overline{\Phi_+} e^{2gV} \Phi_+ + \frac{1}{2} v^2e^{-2gV} + g v^2  V) \nonumber\\
& + \frac{1}{2} \delta^2 + \xi \delta - \frac{1}{2} \delta gv^2 
\end{eqnarray}
The equations of motion reads then
\begin{eqnarray}
0 =& \Delta + 2 g \overline{\Phi_+} e^{2gV} \Phi_+  + gv^2 (1- e^{-2gV}) 
\end{eqnarray}
which has the solution
\begin{eqnarray}
e^{-2gV} =& \frac{1}{-2gv^2} \bigg[ -gv^2 - \Delta \pm \sqrt{(gv^2 +   \Delta)^2 + 8 g^2 v^2 |\Phi_+|^2} \bigg] \nonumber\\
=& \frac{(gv^2 +   \Delta)}{2gv^2} \bigg[ 2 + \frac{4g^2 v^2 |\Phi_+|^2}{(gv^2 +   \Delta)^2} + ...\bigg] 
\end{eqnarray}
Neglecting the terms with derivatives (i.e. $\Delta$), this leads
\begin{eqnarray}
gV =& - \frac{|\Phi_+|^2}{v^2} + 3\frac{|\Phi_+|^4}{v^4} + ...
\end{eqnarray}
which put back into the action, gives:
\begin{eqnarray}
\mathcal{L} =& \int d^2 \theta \frac{1}{\sqrt{2}} mv\Phi_ + + h.c. + \int d^4 \theta \, \, \overline{\Phi_+}  \Phi_+ \bigg[ 1 - \frac{m^2}{2g^2 v^2} \bigg] + |\overline{\Phi_+}  \Phi_+|^2 \bigg[ - \frac{1}{v^2} + \frac{3 m^2}{g^2 v^4} \bigg] + ... 
\label{chiraldominantLagrange}
\end{eqnarray}
As discussed for instance in \cite{Komargodski:2009rz},  (\ref{chiraldominantLagrange})  leads to the nilpotency $\Phi_+$ .

\subsubsection*{Case $\mathbf{m^2\gg g^2v^2}$:}
In this limit, he goldstino is mainly the gaugino $\lambda$. The equation of motion for $\Phi_+$:
\begin{equation} 0 = -4\sqrt{2}mv + D^2(e^{2gV}\Phi_+) + \overline{D}^2(\overline{\Phi_+}e^{2gV}).\end{equation}
has one obvious solution at low energy:
\begin{equation} \Phi_+ = c X_{NL}, \qquad V = \overline{X_{NL}}X_{NL}/\Lambda^2,\end{equation}
in which $c$  and $\Lambda$ can be determined from the vev of the auxiliary field of $\Phi_+$, $V$ and $X_{NL}$.

\subsection{Nilpotent chiral superfield from Ferrara-Zumino supercurrent}
It was conjuctured in \cite{Komargodski:2009rz} that the nilpotent goldstino superfield  controls the non-conservation of the Ferrara-Zumino supercurrent $\mathcal{J}_{\alpha \dot{\alpha}}$ :
\begin{eqnarray}
\overline{D}^ {\dot{\alpha}} \mathcal{J}_{\alpha \dot{\alpha}} = D_\alpha X
\end{eqnarray}

Subsequently  \cite{Arnold:2012yi} have shown that in the presence of a FI term, $X$ can be formally obtained \emph{ in a gauge invariant form} as :
\begin{eqnarray}
X = 4 W - \frac{1}{3} \overline{D}^2 \left[ K + 2 \xi (V + i \Lambda - i \Lambda^\dagger ) \right].\label{FZX}
\end{eqnarray}
For the Lagrangian (\ref{Eq:FILagrangeSU}):
\begin{eqnarray} W &&= \frac{1}{\sqrt{2}} mv\Phi_+,\\
K &&= \overline{\Phi_+}e^{2gV}\Phi_+ + \frac{1}{2}v^2e^{-2gV}.
\end{eqnarray}
The eq. (\ref{FZX}) gives then:
\begin{eqnarray} 
X && = \frac{4\sqrt{2}}{3}mv\Phi_+ - \frac{2}{3}\frac{m^2}{g}\overline{D}^2V + ..., 
\label{XUV}
\end{eqnarray}
in which ... contain higher dimension operator than single fermion. The $\theta$ component of eq. (\ref{FZX}) reads:
\begin{equation} 
X|_{\theta} = \frac{8}{3}mv\psi_+ + \frac{8}{3}\frac{m^2}{g}\lambda.
\label{X|theta}
\end{equation}
 In the IR, the low energy Volkov-Akulov Lagrangian takes the form:
\begin{equation} \mathcal{L}_{VA} = \int d^4\theta \overline{X_{NL}}X_{NL} + (\int d^2\theta -\tilde{f}X_{NL} + h.c.), \qquad \tilde{f} = -\frac{m}{\sqrt{2}g}\sqrt{m^2 + g^2v^2}
\end{equation}
in which the nilpotent chiral superfield $X_{NL}$ :
\begin{equation} 
X_{NL} = \frac{\tilde{\lambda}\tilde{\lambda}}{2\tilde{f}} + \sqrt{2}\theta\tilde{\lambda} + \theta\theta \tilde{f},
\end{equation}
which allows us to identify:
\begin{equation} X = -\frac{8\tilde{f}}{3}X_{NL}.\label{XIR}\end{equation}
By matching eq. (\ref{XUV}) and (\ref{XIR}) we obtain:
\begin{equation} 2\sqrt{2}mv\Phi_+ - \frac{1}{3}\overline{D}^2 (\overline\Phi_+ e^{2gV}\Phi_+ + \frac{1}{2}v^2e^{-2gV} + 2\xi V) \rightarrow  -\frac{8\tilde{f}}{3}X_{NL}.\end{equation}
The $\theta$ component gives :
\begin{equation} \tilde{\lambda} = \frac{gv\psi_+ + m\lambda}{\sqrt{m^2 + g^2v^2}}. \end{equation}
while the auxiliary fields of $\Phi_+$ and $V$ fix:
\begin{eqnarray} \Phi_+ && \rightarrow \frac{gv}{\sqrt{m^2 + g^2v^2}}X_{NL}\\
V && \rightarrow -\frac{g}{m^2 + g^2v^2}\overline{X_{NL}}X_{NL}
\end{eqnarray}

 In summary, for the FI model in the  region of parameters considered here, we can write:
\begin{equation} W_{NL}^\alpha = \frac{1}{4\sqrt{2}f}\overline{D}^2D^\alpha(\overline{X_{NL}}X_{NL}),
\end{equation}
where $f = |F_{X_{NL}}|$. In components, this reads:

\begin{eqnarray} W_{NL}^\alpha = &&\tilde{\lambda}_\alpha + \theta_\alpha\tilde{D} + \frac{i}{2} (\sigma^\mu\overline{\sigma}^\nu\theta)_\alpha \tilde{F}_{\mu\nu} + i\theta\theta (\sigma^\mu\partial_\mu\overline{\tilde{\lambda}})_\alpha,
\end{eqnarray}
where:
\begin{eqnarray} 
\tilde{\lambda}_\alpha &&= -\frac{\overline{F_X}}{f}\psi_\alpha - \frac{i}{f}\partial_\mu\phi (\sigma^\mu\overline{\psi})_\alpha\\
\tilde{D} &&= -\frac{\sqrt{2}F_X\overline{F_X}}{f} + \sqrt{2}\partial^\mu\overline{\phi}\partial_\mu\phi - \frac{i}{\sqrt{2}}\overline{\psi}\overline{\sigma^\mu}\partial_\mu\psi - \frac{i}{\sqrt{2}} \psi\sigma^\mu\partial_\mu\overline{\psi}\\
\tilde{A^\mu} &&= -\frac{\psi\sigma^\mu\overline{\psi} + i\overline{\phi}\partial^\mu\phi - i\phi\partial^\mu\overline{\phi}}{\sqrt{2}f}.
\end{eqnarray}

Using $X_{NL}^2 = 0$ and ${X_{NL}} D^\alpha X_{NL} = 0$, we can show that $W_{NL}^\alpha$ satisfies
\begin{equation} 
\overline{X_{NL}}X_{NL}W_{NL}^\alpha = 0.
\end{equation}

\section{Non-linear Supersymmetry for the Minimal Model of Gravitino Dark Matter}

 Pushing up the scale of supersymmetry breaking leads to a peculiar cosmological scenario where supersymmetry is broken at a scale above the reheating temperature \cite{Benakli:2017whb}. The goldstino as discussed above is "eaten" by the gravitino and gets a $\MP$ suppressed mass. We assume that the soft-terms of sparticles are suppressed  by a messenger scale $\Lambda_{mess}$:
\begin{equation}
\MG= \frac {F}{\sqrt{3} \MP}, \qquad \Ms = \frac {F}{\Lambda_{mess}} .
\end{equation}
and that $m_{3/2} \ll M_{SUSY}$ as a consequence of ${\Lambda_{mess}}\ll \MP$ which is the case for gauge mediation. The low energy particle content consists then only in Standard Model states and a gravitino.  Our cosmological history described here starts after the Universe is reheated. Some assumptions are made for this epoch: (i)  The 
reheating temperature $T_{RH}$ is small enough to not produce superpartners of the Standard Model particles, thereof $T_{RH}
 \lesssim \Ms$ (ii) in the reheating process gravitinos are scarcely produced (for other possibilities, see \cite{Dudas:2017rpa}). The whole spectrum then goes as follows:
\begin{equation}
\MG \ll  T_{RH} \lesssim \Ms \lesssim \sqrt{F} \lesssim {\Lambda_{mess}}\ll \MP 
\end{equation}
When the energy energy scale $E$ of the gravitinos is much bigger than their mass, the dark matter gravitino interactions are well approximated by the helicity $\pm 1/2$ component, the goldstino in virtue of the equivalence theorem \cite{Fayet:1977vd}. Leading order goldstino-matter interactions corresponding to the minimal couplings expected from the low energy 
 theorem read: 
\begin{equation}
L_{2G} = \frac{i}{2F^2}(G\sigma^\mu\partial^\nu\bar{G} - \partial^\nu G\sigma^\mu\bar{G}) T_{\mu\nu},
\end{equation}
where $G$ is the goldstino field and $T_{\mu\nu}$ is the energy momentum tensor of the SM matter fields. As $\MG$ is much smaller than $T_{RH}$, the $2\rightarrow2$ scatterings for the goldstino production is dominated by the dimension eight operators coupled with the kinetic terms
\begin{eqnarray}
&&
\frac{i}{2F^2} (G\sigma^\mu\partial^\nu\bar{G} - \partial^\nu G \sigma^\mu  \bar{G}) ( \partial_\mu H \partial_\nu H^\dagger +
\partial_\nu H \partial_\mu H^\dagger) , \nonumber
\\
&&
\frac{1}{8F^2} (G\sigma^\mu\partial^\nu\bar{G} - \partial^\nu G \sigma^\mu \bar{G} )  \times  \nonumber
\\
&&
 (\bar{\psi}\bar{\sigma}_\nu \partial_\mu\psi + \bar{\psi} \bar{\sigma}_\mu \partial_\nu\psi 
 -\partial_\mu {\psi} \bar{\sigma}_\nu \psi -   \partial_\nu {\psi} \bar{\sigma}_\mu \psi ) ,
 \nonumber
 \\
&&
\sum_{a} \frac{i}{2F^2} (G\sigma^\xi\partial_\mu\bar{G} -  \partial_\mu G \sigma^\xi  \bar{G} ) F^{\mu\nu a}F^a_{\nu\xi},
\label{Eq:gprod}
\end{eqnarray}
\noindent
where $h$, $\psi$ and $F^a_{\nu\xi}$ stand for a complex scalar (Higgs doublet), gauge bosons and two-component fermions (quarks and leptons), respectively. 
\noindent
From the interaction generated through the Lagrangian Eq.(\ref{Eq:gprod}), one can compute
 the production rate $R = n_{eq}^2 \langle \sigma v \rangle$ 
 of the gravitino $\tilde G$, generated  by the annihilation of the standard model bath of density
 $n_{eq}$: 
 \begin{equation}
R = \sum_i n_{eq}^2 \langle \sigma v \rangle_i \simeq 21.65 \times \frac{T^{12}}{F^4}.
\label{Eq:Rfinal}
\end{equation}
Thus the contribution to the yields is dominant at the temperature near $T_{RH}$.  The gravitino relic abundance is:
 \begin{equation}
 \Omega_{3/2}h^2 \simeq 
 0.11 \left( \frac{100 ~\mrm{GeV}}{m_{3/2}} \right)^3 \left( \frac{T_{\mrm{RH}}}{5.4 \times10^7 ~\mrm{GeV}} \right)^7
 \label{Eq:omega}
  \end{equation}
As we notice, the dependence on the reheating temperature is completely different from  standard gravitino dark matter scenario. Due to the 
  large power $T^7_{\mrm{RH}}$, details as the total number of degrees of freedom, or even channels, does not influence that much the final reheating temperature.

\vskip.1in
\noindent
{\bf Acknowledgments}

\noindent We are grateful to I. Antoniadis, J. P. Derendinger,E. Dudas and M. Goodsell for useful discussions.  This work  is supported by the Labex ``Institut Lagrange de Paris'' (ANR-11-IDEX-0004-02,  ANR-10-LABX-63) and by the Agence Nationale de Recherche under grant ANR-15-CE31-0002 ``HiggsAutomator''. K.B acknowledges also the support of the the European Research Council (ERC) under the Advanced  Grant Higgs@LHC (ERC-2012-ADG20120216-321133).



\begin{thebibliography} {99}

\bibitem{Volkov:1973ix} 
  D.~V.~Volkov and V.~P.~Akulov,
  ``Is the Neutrino a Goldstone Particle?,''
  Phys.\ Lett.\  {\bf 46B}, 109 (1973).
  
\bibitem{Rocek:1978nb}
  M.~Ro\v{c}ek,
  ``Linearizing the Volkov-Akulov Model,''
  Phys.\ Rev.\ Lett.\  {\bf 41} (1978) 451.

\bibitem{Benakli:2017yar}
  K.~Benakli, Y.~Chen and M.~D.~Goodsell,
  arXiv:1711.08466 [hep-th].
  
\bibitem{Casalbuoni:1988xh}
  R.~Casalbuoni, S.~De Curtis, D.~Dominici, F.~Feruglio and R.~Gatto,
  ``Nonlinear Realization of Supersymmetry Algebra From Supersymmetric Constraint,''
  Phys.\ Lett.\ B {\bf 220} (1989) 569.


\bibitem{Komargodski:2009rz}
  Z.~Komargodski and N.~Seiberg,
  ``From Linear SUSY to Constrained Superfields,''
  JHEP {\bf 0909} (2009) 066
  [arXiv:0907.2441 [hep-th]].
  
\bibitem{Lindstrom:1979kq}
  U.~Lindstrom and M.~Ro\v{c}ek,
  ``Constrained Local Superfields,''
  Phys.\ Rev.\ D {\bf 19} (1979) 2300.

\bibitem{Ivanov:1978mx}
  E.~A.~Ivanov and A.~A.~Kapustnikov,
  J.\ Phys.\ A {\bf 11} (1978) 2375.
  doi:10.1088/0305-4470/11/12/005
  
\bibitem{Ivanov:1982bpa}
  E.~A.~Ivanov and A.~A.~Kapustnikov,
  J.\ Phys.\ G {\bf 8} (1982) 167.
  doi:10.1088/0305-4616/8/2/004
  
\bibitem{Samuel:1982uh}
  S.~Samuel and J.~Wess,
  ``A Superfield Formulation of the Nonlinear Realization of Supersymmetry and Its Coupling to Supergravity,''
  Nucl.\ Phys.\ B {\bf 221} (1983) 153.


\bibitem{Bandos:2016xyu}
  I.~Bandos, M.~Heller, S.~M.~Kuzenko, L.~Martucci and D.~Sorokin,
  JHEP {\bf 1611} (2016) 109
  doi:10.1007/JHEP11(2016)109
  [arXiv:1608.05908 [hep-th]].
  
\bibitem{Cribiori:2017ngp} 
  N.~Cribiori, G.~Dall'Agata and F.~Farakos,
  ``From Linear to Non-linear SUSY and Back Again,''
  JHEP {\bf 1708}, 117 (2017)
  [arXiv:1704.07387 [hep-th]].

\bibitem{Buchbinder:2017qls}
  E.~I.~Buchbinder, J.~Hutomo, S.~M.~Kuzenko and G.~Tartaglino-Mazzucchelli,
  Phys.\ Rev.\ D {\bf 96} (2017) no.12,  126015
  doi:10.1103/PhysRevD.96.126015
  [arXiv:1710.00554 [hep-th]].
  
\bibitem{GarciadelMoral:2017vnz} 
  M.~P.~Garcia del Moral, S.~Parameswaran, N.~Quiroz and I.~Zavala,
  ``Anti-D3 branes and moduli in non-linear supergravity,''
  JHEP {\bf 1710}, 185 (2017)
  [arXiv:1707.07059 [hep-th]].
  

\bibitem{Luo:2009ib}
  H.~Luo, M.~Luo and S.~Zheng,
  ``Constrained Superfields and Standard Realization of Nonlinear Supersymmetry,''
  JHEP {\bf 1001} (2010) 043
  [arXiv:0910.2110 [hep-th]].
  
\bibitem{Liu:2010sk}
  H.~Liu, H.~Luo, M.~Luo and L.~Wang,
  ``Leading Order Actions of Goldstino Fields,''
  Eur.\ Phys.\ J.\ C {\bf 71} (2011) 1793
  [arXiv:1005.0231 [hep-th]].
  
\bibitem{Kuzenko:2010ef}
  S.~M.~Kuzenko and S.~J.~Tyler,
  ``Relating the Komargodski-Seiberg and Akulov-Volkov actions: Exact nonlinear field redefinition,''
  Phys.\ Lett.\ B {\bf 698} (2011) 319
  [arXiv:1009.3298 [hep-th]].
  
\bibitem{Ferrara:2015tyn}
  S.~Ferrara, R.~Kallosh and J.~Thaler,
  ``Cosmology with orthogonal nilpotent superfields,''
  Phys.\ Rev.\ D {\bf 93} (2016) no.4,  043516
  [arXiv:1512.00545 [hep-th]].
  
\bibitem{Dalianis:2017okk} 
  I.~Dalianis and F.~Farakos,
  ``Constrained superfields from inflation to reheating,''
  Phys.\ Lett.\ B {\bf 773}, 610 (2017)
  [arXiv:1705.06717 [hep-th]].
  
  
\bibitem{DallAgata:2016syy}
  G.~Dall'Agata, E.~Dudas and F.~Farakos,
  ``On the origin of constrained superfields,''
  JHEP {\bf 1605} (2016) 041
  [arXiv:1603.03416 [hep-th]].
  
 
\bibitem{Fayet:1974jb}
  P.~Fayet and J.~Iliopoulos,
  ``Spontaneously Broken Supergauge Symmetries and Goldstone Spinors,''
  Phys.\ Lett.\  {\bf 51B} (1974) 461.
  
\bibitem{Arnold:2012yi} 
  D.~Arnold, J.~P.~Derendinger and J.~Hartong,
  ``On Supercurrent Superfields and Fayet-Iliopoulos Terms in N=1 Gauge Theories,''
  Nucl.\ Phys.\ B {\bf 867}, 370 (2013)
  [arXiv:1208.1648 [hep-th]].
  
\bibitem{Goodsell:2014dia}
  M.~D.~Goodsell and P.~Tziveloglou,
  ``Dirac Gauginos in Low Scale Supersymmetry Breaking,''
  Nucl.\ Phys.\ B {\bf 889} (2014) 650
  [arXiv:1407.5076 [hep-ph]].

\bibitem{Nelson:2015cea}
  A.~E.~Nelson and T.~S.~Roy,
  ``New Supersoft Supersymmetry Breaking Operators and a Solution to the $\mu$ Problem,''
  Phys.\ Rev.\ Lett.\  {\bf 114} (2015) 201802
  [arXiv:1501.03251 [hep-ph]].
 
\bibitem{Antoniadis:2010hs}
  I.~Antoniadis, E.~Dudas, D.~M.~Ghilencea and P.~Tziveloglou,
  ``Non-linear MSSM,''
  Nucl.\ Phys.\ B {\bf 841} (2010) 157
  [arXiv:1006.1662 [hep-ph]].
  
 
\bibitem{Benakli:2017whb}
  K.~Benakli, Y.~Chen, E.~Dudas and Y.~Mambrini,
  Phys.\ Rev.\ D {\bf 95} (2017) no.9,  095002
  [arXiv:1701.06574 [hep-ph]].

  
\bibitem{Dudas:2017rpa} 
  E.~Dudas, Y.~Mambrini and K.~Olive,
  Phys.\ Rev.\ Lett.\  {\bf 119}, no. 5, 051801 (2017)
  [arXiv:1704.03008 [hep-ph]].
  M.~A.~G.~Garcia, Y.~Mambrini, K.~A.~Olive and M.~Peloso,
  Phys.\ Rev.\ D {\bf 96}, no. 10, 103510 (2017)
  [arXiv:1709.01549 [hep-ph]].
  E.~Dudas, T.~Gherghetta, Y.~Mambrini and K.~A.~Olive,
  Phys.\ Rev.\ D {\bf 96}, no. 11, 115032 (2017)
  [arXiv:1710.07341 [hep-ph]].
  
    
  
  \bibitem{Fayet:1977vd}
  P.~Fayet,
  Phys.\ Lett.\  {\bf 70B} (1977) 461;
  P.~Fayet,
  Phys.\ Lett.\  {\bf 86B} (1979) 272.

\end{thebibliography}
\end{document}